\documentclass[conference]{IEEEtran}
\IEEEoverridecommandlockouts
\usepackage{cite}
\usepackage{amsmath,amssymb,amsfonts}
\usepackage{algorithmic}
\usepackage{graphicx}
\usepackage{textcomp}
\usepackage{xcolor}

\usepackage[english]{babel}

\usepackage{pstool}                    
\usepackage[dvips]{rotating,graphicx}  
\usepackage{float}                     
\usepackage{examplep}

\usepackage{hyperref}

\usepackage{enumitem}  
\usepackage{latexsym}
\usepackage[english]{babel}
\usepackage{csquotes}

\usepackage{siunitx}                        
\usepackage{amsmath}                        
\usepackage{mathtools}                      
\usepackage{bm}  

\usepackage{multirow}       
\usepackage{booktabs}       
\usepackage{makecell}       
\usepackage{tikz}
\usepackage{ragged2e}
\usepackage{xcolor} 
\usepackage{graphicx} 
\usepackage{cite}
\usepackage[nameinlink,capitalize]{cleveref}  

\usepackage{url} 

\def\BibTeX{{\rm B\kern-.05em{\sc i\kern-.025em b}\kern-.08em
    T\kern-.1667em\lower.7ex\hbox{E}\kern-.125emX}}
\begin{document}

\title{Multi-GPU-Enabled Hybrid Quantum-Classical Workflow in Quantum-HPC Middleware:\\ Applications in Quantum Simulations

\thanks{The first two authors contributed equally to this work. }
}

\author{
    \IEEEauthorblockN{
        Kuan-Cheng Chen\IEEEauthorrefmark{1}\IEEEauthorrefmark{2}\IEEEauthorrefmark{6},
        Xiaoren Li\IEEEauthorrefmark{3},
        Xiaotian Xu\IEEEauthorrefmark{1}\IEEEauthorrefmark{2},
        Yun-Yuan Wang\IEEEauthorrefmark{4}
        Chen-Yu Liu\IEEEauthorrefmark{5},
    }
    \IEEEauthorblockA{
        \IEEEauthorrefmark{1}Centre for Quantum Engineering, Science and Technology (QuEST), Imperial College London, SW7 2BX London, UK\\
        \IEEEauthorrefmark{2}Department of Materials, Imperial College London, SW7 2BX, London, UK
    }
    \IEEEauthorblockA{
        \IEEEauthorrefmark{3}Department of Physics and Astronomy, University of Waterloo, ON N2L 3G1 Waterloo, Canada
    }
     \IEEEauthorblockA{
        \IEEEauthorrefmark{4}NVidia AI Technology Center (NVAITC), NVIDIA Corp., 114 Taipei, Taiwan
    }
    \IEEEauthorblockA{
        \IEEEauthorrefmark{5}Graduate Institute of Applied Physics, National Taiwan University, 10663 Taipei, Taiwan
    }

    \IEEEauthorblockA{
        \IEEEauthorrefmark{6}Email: kuan-cheng.chen17@ic.ac.uk
    }
}

\maketitle

\begin{abstract}
Achieving high-performance computation on quantum systems presents a formidable challenge that necessitates bridging the capabilities between quantum hardware and classical computing resources. This study introduces an innovative distribution-aware Quantum-Classical-Quantum (QCQ) architecture, which integrates cutting-edge quantum software frameworks with high-performance classical computing resources to address challenges in quantum simulation for materials and condensed matter physics, including the prediction of quantum phase transitions. At the heart of this architecture is the seamless integration of Variational Quantum Eigensolver (VQE) algorithms running on Quantum Processing Units (QPUs) for efficient quantum state preparation, Tensor Network states, and Quantum Convolutional Neural Networks (QCNNs) for classifying quantum states on classical hardware.

For benchmarking quantum simulators, the QCQ architecture utilizes the cuQuantum SDK to leverage multi-GPU acceleration, integrated with PennyLane's Lightning plugin, demonstrating up to tenfold increases in computational speed for complex phase transition classification tasks compared to traditional CPU-based methods. This significant acceleration enables models such as the transverse field Ising and XXZ systems to accurately predict phase transitions with a 99.5\% accuracy. The architecture's ability to distribute computation between QPUs and classical resources addresses critical bottlenecks in quantum High-Performance Computing (HPC), paving the way for scalable quantum simulation.

The QCQ framework embodies a synergistic combination of quantum algorithms, machine learning, and Quantum-HPC capabilities, enhancing its potential to provide transformative insights into the behavior of quantum systems across different scales. As quantum hardware continues to improve, this hybrid distribution-aware framework will play a crucial role in realizing the full potential of quantum computing by seamlessly integrating distributed quantum resources with the state-of-the-art classical computing infrastructure.
\end{abstract}


\section{Introduction}

Quantum Machine Learning (QML) integrates the disciplines of machine learning and quantum computing\cite{biamonte2017quantum}, employing parameterized quantum circuits as statistical models\cite{benedetti2019parameterized}. This technology has seen an increasing array of applications in the natural sciences\cite{ngairangbam2022anomaly}, generative modelling\cite{zoufal2019quantum}, and classification problems\cite{mari2020transfer,chen2022quantum}. Due to its high expressivity\cite{abbas2021power}, QML demonstrates superior performance over conventional models across numerous domains. This includes stellar classification within large datasets\cite{chen2023quantum}, leveraging an architecture that effectively bridges classical data with quantum algorithms. However, current applications are constrained by the limited number of Quantum Processing Unit (QPU) qubits and the fidelity of qubits in the Noisy Intermediate-Scale Quantum (NISQ) era\cite{marshall2023high}. Despite these challenges, recent research conducted by Q-CTRL and IBM Quantum has made progress in enhancing the success probability of algorithms through an automated deterministic error-suppression workflow and quantum error mitigation technique\cite{mundada2023experimental, chen2023short}. Nonetheless, the availability of qubits remains a significant issue in the design of quantum machine learning algorithms\cite{marshall2023high}.

On the other hand, the work of Huang et al. has demonstrated quantum advantage through the utilization of quantum data with a quantum computer\cite{huang2022quantum}. For instance, Monaco et al.'s application of a Variational Quantum Eigensolver (VQE)-based quantum neural network model for quantum phase detection in the axial next-nearest-neighbour Ising (ANNNI) model illustrates its superiority\cite{monaco2023quantum}. However, such architectures necessitate an increased number of qubits for learning more complex models\cite{grzesiak2020efficient}. The demand for a large number of qubits can be addressed by integrating architectures through distributed quantum computing\cite{cuomo2020towards}, also referred to as quantum-centric supercomputing by IBM Quantum\cite{alexeev2023quantum}. In the realm of compact quantum devices, achieving high-fidelity qubit operations and their subsequent verification is relatively manageable. Nonetheless, the shift towards modular approaches mandates the transfer of quantum information between QPUs, thereby fostering effective qubit interactions across various modules. While this approach offers certain benefits, the ensuing inter-module communication is often characterized by slower speeds and reduced reliability, leading to the emergence of a challenge known as the quantum interconnect bottleneck (QIB)\cite{awschalom2021development}. Furthermore, the adoption of distributed quantum algorithms introduces vulnerabilities that necessitate the formulation of robust quantum protocols to ensure system integrity and security\cite{prest2023quantum}.

In this study, we endeavor to develop a scheme that is conducive to high-performance computing (HPC) within the quantum domain \cite{saurabh2023conceptual}, specifically designed to enable distributed quantum computing through a hybrid quantum-classical workflow. This framework incorporates multi-GPU acceleration (via the cuQuantum SDK\cite{cuQuantum}) to support quantum simulators (which serve as counterparts to noiseless QPUs, if QPUs are not available). Additionally, we integrate the latest Pennylane Lightning plugins\cite{bergholm2018pennylane}. This component leverages CUDA-aware MPI (Message Passing Interface) boosted by NVLink's 600 GB/s bidirectional bandwidth for optimized GPU-to-GPU communication, enabling rapid distributed simulation tasks\cite{gao2021scaling}. This middleware scheme offers a synergy between classical GPU acceleration and quantum computing acceleration\cite{wintersperger2022qpu, saurabh2023conceptual}. The relevant conceptual architectures for Quantum-HPC middleware are discussed by Saurabh et. al\cite{saurabh2023conceptual}. 

Echoing the architectural paradigm introduced by Monaco et al.\cite{monaco2023quantum}, our methodology adopts a VQE-based ansatz to explore quantum phase transitions in both the transverse field Ising and the XXZ models\cite{uvarov2020machine}. Different from the quantum-to-quantum or the quantum-to-classical architecture\cite{monaco2023quantum,uvarov2020machine,khait2023variational,liu2023learning}, we propose an quantum-classical-quantum sandwich (QSandwich) architecture. This architecture capitalizes on the capabilities of classical convolutional neural networks (CNNs) to significantly reduce the qubit requirements for the quantum classifier. 

For the implementation of the VQE-based ansatz, we also explore the potential of distributed-VQE approaches\cite{diadamo2021distributed,du2022distributed,khait2023variational} or the employment of multi-GPU configurations\cite{modani2023multi} to meet the demands of large qubit requirements in complex scenarios. The QSandwich framework is designed to ensure that each quantum layer is effectively managed and utilized within a distributed quantum computing environment, leveraging small, yet high-fidelity QPU resources. This approach underscores our commitment to advancing the frontier of quantum computing by optimizing computational resources and fidelity in the execution of quantum simulations.

\section{Applications and Algorithms}

\subsection{QSandwitch: Quantum-Classical-Quantum framework for Quantum-HPC Processing}

\begin{figure}[htpb]
    \centering
    \includegraphics[width=0.5\textwidth]{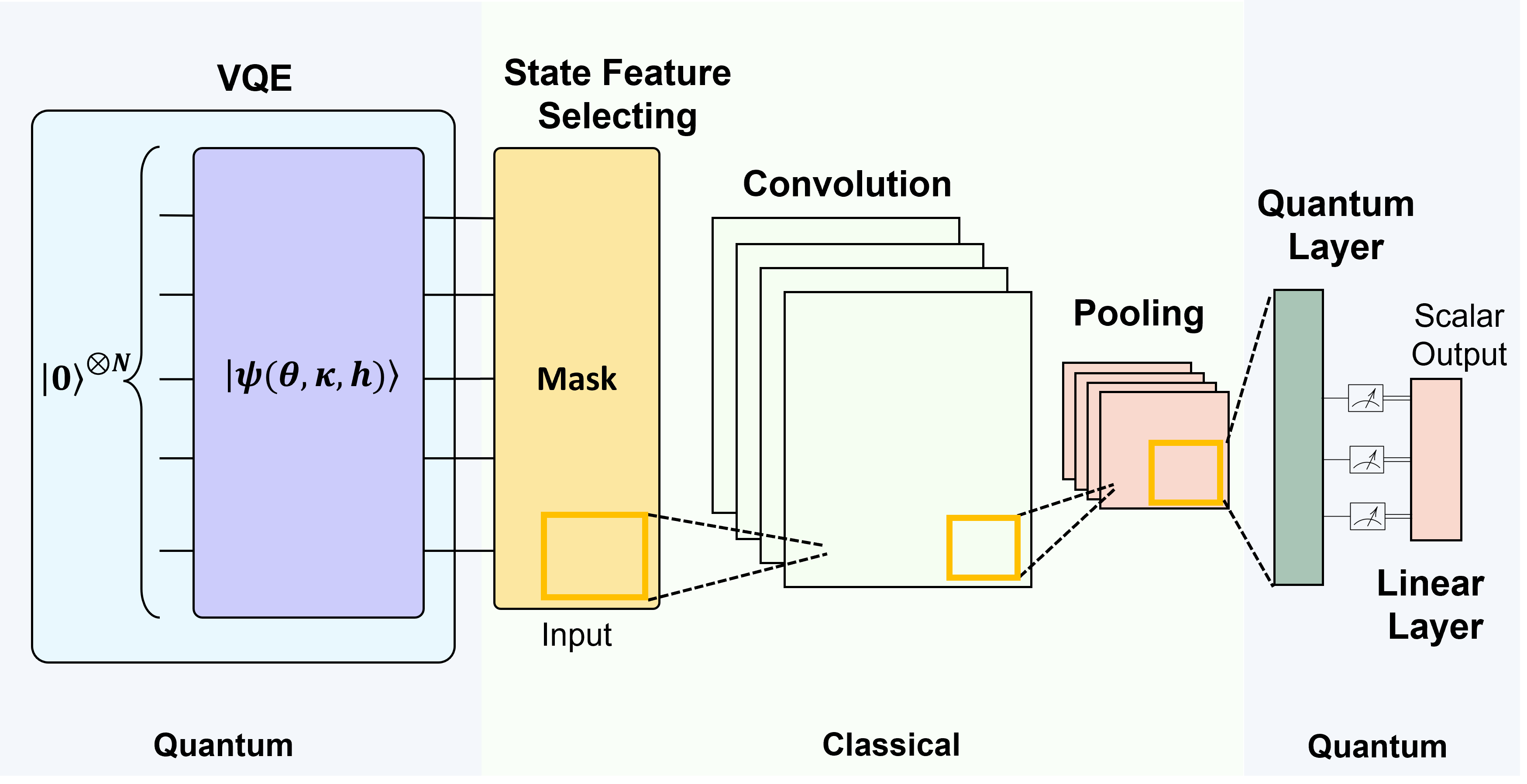}
    \caption{Pipeline of the Hybrid QSandwich Architecture.}
    \label{fig:QCQ}
\end{figure}

In the domain of Quantum-HPC\cite{saurabh2023conceptual}, we proposed a novel distribution-aware hybrid quantum-classical-quantum (QCQ) processing framework shown in Fig. \ref{fig:QCQ}, representing an optimized approach to the computational challenges of distributed quantum computing for classifying phase transition. Initiated by the VQE algorithm, this framework begins its operation within the quantum realm, where a specific quantum state is prepared to encode potential solutions. This quantum state is then converted into classical information via a process termed state feature selection, during which essential characteristics of the quantum state are identified and extracted for subsequent processing in classical CNN layers. This methodology introduces a structured integration of quantum and classical computing techniques, aiming to optimize the solution-finding process through the strategic manipulation and analysis of quantum data.

The classical segment of the QCQ framework harnesses the capabilities of CNNs to scrutinize the quantum state prepared by the VQE algorithm. This involves the application of convolutional layers to refine the data, augmented by pooling layers that distil the essence of the information. The refined data is then adeptly re-encoded into quantum form, traversing another quantum layer tasked with extracting a scalar output to classify the phase transition problem.

The scalar output generated by the process is not the end point but rather an intermediate stage requiring additional refinement via a sigmoid layer to improve the accuracy of the final result. The following section will provide a detailed explanation of this hybrid QCQ architecture, focusing on the interaction between its quantum and classical components. The integration of these computational realms aims to enhance the capabilities of the near-term Quantum-HPC ecosystem, addressing challenges with efficiency beyond what quantum or classical computing can achieve individually.

\subsubsection{VQE for State Preparation}
\begin{figure}[htpb]
\centering
\includegraphics[width=0.5\textwidth]{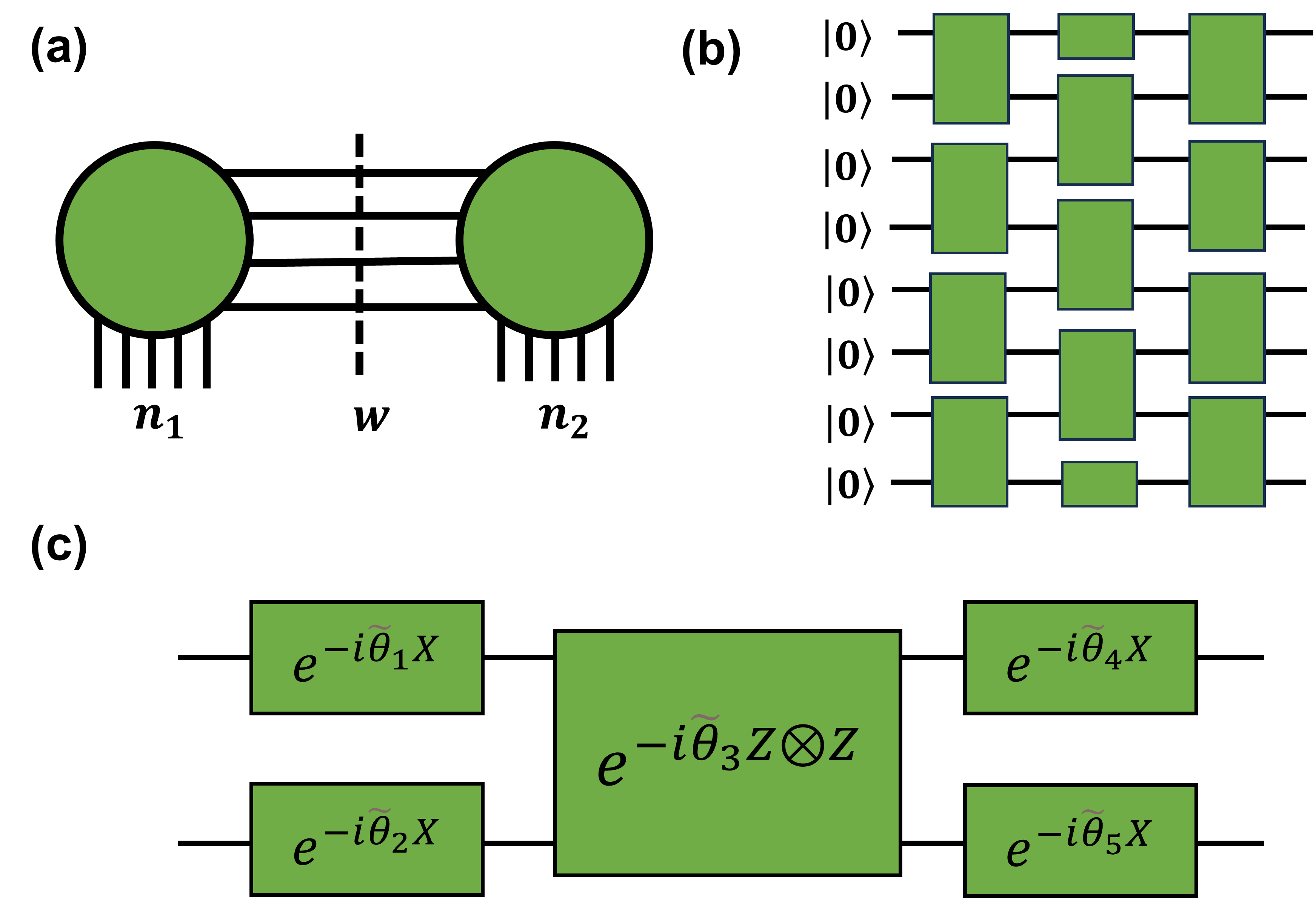}
\caption{(a) Quantum circuit depicted as a tensor network with bonds of dimension 2. (b) Checkerboard tensor network circuit. \cite{cuQuantum} Each green block refers to a two-qubit entangler circuit. (c) The entangled block in the checkerboard circuit}
\label{fig: tensor}
\end{figure}

VQE is a common approach in ground state preparation in quantum computing\cite{liu2019variational}. In our study, we focus on preparing the ground state for the phase-transition problem with VQE and tensor network ansatz\cite{uvarov2020machine}. These methods enable efficient and accurate approximation of quantum states. Here are the key points of our approach:
\begin{itemize}

    \item \textbf{Ground State Preparation}: 
    We employ the VQE to approximate the ground states of given quantum Hamiltonians. The advantage of using VQE for state preparation, instead of loading state information from the Pennylane dataset\cite{Utkarsh2023Spin}, is that, by knowing the circuit structure, we can reproduce the ground state completely. On the other hand, the classical-shadow information from the Pennylane dataset will be helpful to reproduce the state more accurately\cite{huang2020predicting}, however, it could not be the same state.
    Knowing the circuit structure also reduces the size of datasets which speeds up the data reading process in quantum machine learning later on. A n-qubits states information from pennylane datasets contains $2^n$ floats. On the other hand, our VQE states preparation datasets only records the parameters for the ansatz circuits. The number of parameters varies depending on the structure and depth of the ansatz. For example, we used 100 float parameters for a 10-qubit circuit in state preparation which reduced the size of datasets (otherwise 1024 floats needed) and performance well in classification.

    Additionally, building the state preparation ourselves allows us to prepare more ground states (100 states from the Pennylane dataset\cite{Utkarsh2023Spin} and 1000 states from our preparation) and makes it possible to apply data augmentation.

    \item \textbf{Tensor Network Ansatz}:  Our approach replaces the standard unitary coupled cluster ansatz with tensor network ansatz states introduced by Uvarov et. al\cite{uvarov2020machine}, which was inspired by tensor networks, to prepare the ground states efficiently. There are several families of variational ansatz states, including rank-one circuits, tree tensor network circuits, and checkerboard-shaped circuits with varying depths. The structure of the checkerboard-shaped circuit is shown in Fig. \ref{fig: tensor} (b). These ansatz circuits are designed to capture different amounts of entanglement and are critical for accurately approximating the ground states of various Hamiltonians. 
    We choose the checkerboard-shaped circuits in our VQE due to the best performance in reducing the error of ground state energy\cite{uvarov2020machine}. For each entangled blocks in a checkerboard-shaped circuit, we used Ising entangle gates shown in Fig. \ref{fig: tensor} (c).

    \item \textbf{Hamiltonian Decomposition}: We represent the Hamiltonian as a sum of tensor products of Pauli operators. This decomposition allows us to estimate individual terms of the Hamiltonian expectation value variational and minimize the energy using a hybrid process. More math details are shown in \ref{secmodels}
    
    \item \textbf{Optimization}: The parameters of the ansatz states are optimized using classical algorithms to minimize the expectation value of the Hamiltonian. This process iterates until the ground state is approximated within the desired accuracy.
\end{itemize}

\subsubsection{Data Augmentation}

To enhance the robustness of our model and improve its accuracy, we apply data augmentation techniques to the quantum states data. This involves applying rotations and spin flips to the VQE-prepared states, creating additional valid data points for training the classifier. This technique exploits the symmetries of the Hamiltonians to generate a more diverse training dataset. 

Our methodology represents an approach to predicting phase transitions in quantum systems by merging the capabilities of quantum simulation and quantum machine learning. Through the strategic use of tensor network ansatz states, a quantum classifier, and the computational power of multi-GPU frameworks facilitated by PennyLane, we achieve significant improvements in speed and scalability, paving the way for advanced studies in quantum phase transitions.

\subsubsection{Quantum Convolutional Neural Network Classifier (Felix)}
\begin{figure}[htpb]
    \centering
    \includegraphics[width=\linewidth]{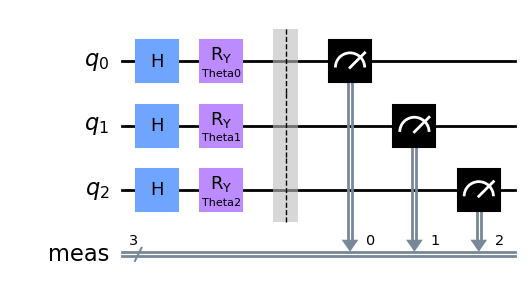}
    \caption{The quantum circuit used in QCNN. In this circuit, a Hadamard gate is first applied to each qubit, followed by a rotational-Y gate. The angles of the rotational gates are the trainable parameters (\texttt{Theta1}, \texttt{Theta2} and \texttt{Theta3}). At the end of this circuit, measurement is executed and the possibilities of finding 7 out of 8 quantum states (except ``111") are the output of this circuit. Thus, when regarded as a layer in the QCNN, this circuit has 3 input channels and 7 output channels.}
    \label{fig: NN circuit}
\end{figure}

Quantum Convolutional Neural Network (QCNN) is utilized to predict phase transitions, harnessing the combined strengths of convolutional layers and quantum computing layers \cite{cong2019quantum}. This approach allows for efficient feature extraction from extensive databases and the natural simulation of quantum data. The QCNN algorithm, as utilized in the QCQ architecture, incorporates a single convolutional layer connected with a pooling layer, multiple fully connected layers, and a quantum circuit layer. This architecture is illustrated in Fig. \ref{fig:QCQ} and is explained below:

\begin{itemize}
    \item \textbf{Convolutional Layer}: The model initiates its computational flow with a single convolutional layer. It utilizes a one-dimensional convolution, generating 30 output channels, with a kernel size identical to that of the feature. This design is specifically tailored for the initial extraction of features directly from the input data.
    \item \textbf{Max Pooling Layer}: One max pooling layer, configured with a kernel size of 1, is placed strategically to reduce spatial dimensions and hence, the complexity of the data passing through the network.
    \item \textbf{Fully Connected Layers}: After flattening the feature map of the max pooling layer, the model incorporates two fully connected layers that serve to interpret the features extracted by the convolutional and pooling layers, culminating in the model's ability to make predictions. Notably, the second fully connected layer is specifically designed to interface with the quantum circuit layer, mapping the classical data into a quantum-compatible format.
    \item \textbf{Quantum Circuit Layer}: At the heart of QCNN approach is the quantum circuit layer, invoked through a predefined Pennylane circuit. The circuit is illustrated in Fig. \ref{fig: NN circuit}. This layer signifies the model's capacity to perform quantum computations, potentially exploiting quantum parallelism and entanglement to enhance the model's predictive capabilities. This quantum layer is followed by another fully connected layer.
    \item \textbf{Dropout}: A dropout layer with a probability (p) of 0.5 is integrated to prevent overfitting by randomly omitting a subset of features during the training phase.
    \item \textbf{Prediction}: The sigmoid activation function is employed to transform the feature map of the final fully connected layer into a probability value ranging between 0 and 1. This mapping facilitates a clear decision-making process for binary classification tasks. The computation of the loss involves measuring the discrepancy between the sigmoid function's output and the designated target values. Upon completion of the training process, the model makes predictions based on a threshold criterion: outputs exceeding 0.5 are classified as 1, indicating the presence of phase transition, while those below 0.5 are classified as 0.
\end{itemize} 

\subsection{Hamiltonians for Solving Phase Transition}\label{secmodels}
In our study, we introduce a novel approach that combines quantum simulation with QML to classify phases of matter, addressing the computational challenges faced by classical simulation methods. By employing a variational quantum algorithm, we leverage the capabilities of quantum computers to prepare and classify labelled states derived from the VQE algorithm. This method effectively bypasses the data reading slowdown typically encountered in quantum-enhanced machine learning applications, presenting a significant advancement in the field.

Our work utilizes families of variational ansatz circuits inspired by tensor networks, enabling us to exploit tensor network theory to elucidate properties of phase diagrams. This approach is instrumental in our development of a quantum neural network. These results underscore the potential of integrating quantum simulation with QML to provide deep computational insights into quantum systems.

We represent the following Hamiltonian as a sum of tensor products of Pauli operators:

\begin{equation}
H = \sum_{\alpha_1\alpha_2...\alpha_n} J_{\alpha_1\alpha_2...\alpha_n} \sigma_{\alpha_1} \otimes \sigma_{\alpha_2} \otimes \cdots \otimes \sigma_{\alpha_n},
\end{equation}

where $\alpha_i \in \{0,1,2,3\}$ enumerate the Pauli matrices $\{1, X, Y, Z\}$. With the decomposition (1), individual terms of $\langle \psi(\theta) | H | \psi(\theta) \rangle$ can be estimated and variationally minimized elementwise using a classical-to-quantum process. In each iteration, one prepares the state $|\psi(\theta)\rangle$ and measures each qubit in the local $X$, $Y$, or $Z$ basis, estimates the energy and updates $\theta$. This method can become scalable only if the number of terms in the Hamiltonian is polynomially bounded in the number of spins and the coefficients $J_{\alpha_1\alpha_2...\alpha_n}$ are defined up to poly(n) digits.

In particular, we use the transverse field Ising model
(TFIM):
\begin{equation}
H_{TFIM} = J\sum_{i}  \sigma_{i}^z\sigma_{i+1}^z+
h\sum_{i}  \sigma_{i}^x, (J>0,h>0)
\end{equation}
where $\sigma_{i}^z$ are Pauli Z matrix acting on the $i$th spin. Constant $J$ is known as the energy scale, we set $J=1$ in our study which means no loss of generality. Dimensionless constant $h$ corresponds to the transverse magnetic field.

When $J>h$, the system has two ground states, with opposite signs
of magnetization.When $J<h$ the system has one ground state with zero magnetization. The changing of ground states refers to the phase changing of the system, which happens at $J=h$.

The other model we used is antiferromagnetic XXZ spin chain model:
\begin{equation}
H_{xxz} = -\sum_{i}J_\perp(\sigma_{i}^x\sigma_{i+1}^x+\sigma_{i}^y\sigma_{i+1}^y)+
 J_z \sigma_{i}^z\sigma_{i+1}^z, (J_z<0)
\end{equation}
where the sign of $J_z$ determines if the system is ferromagnetic ($J_z > 0$) or
antiferromagnetic ($J_z < 0$). The sign of $J_\perp$ is not important in bipartite lattice because the states can be redefined by $\sigma^x \rightarrow -\sigma^x, \sigma^y \rightarrow -\sigma^y$. We set $J_\perp=1$ in our study.

For antiferromagnetic XXZ model, when $|J_z|>|J_\perp|$, the ground states spins each takes their opposite values to their neighbours, and the system is in antiferromagnetic Ising state; When $|J_z|<|J_\perp|$, the system is in the planar phase. A Berezinsky–Kosterlitz–Thouless type phase transition happens at $|J_z|=|J_\perp|$ \cite{xxzphase}.

Conclusively, our research demonstrates the feasibility and effectiveness of using quantum simulation and QML to classify phases of matter. By preparing approximate ground states variationally and employing them as inputs to a quantum classifier, we avoid the limitations of traditional Monte Carlo sampling methods. The success of our nearest-neighbour quantum neural network in accurately identifying phases of matter highlights the promising future of this interdisciplinary approach. This synergy between quantum computing and machine learning opens new avenues for exploring quantum systems, significantly impacting the fields of condensed matter physics and quantum computing.

\subsection{Towrad Distributed Quantum Computing: a Conceptual Architecture for a Distributed-Quantum-HPC Middleware}

\begin{figure}[htpb]
    \centering
    \includegraphics[width=\linewidth]{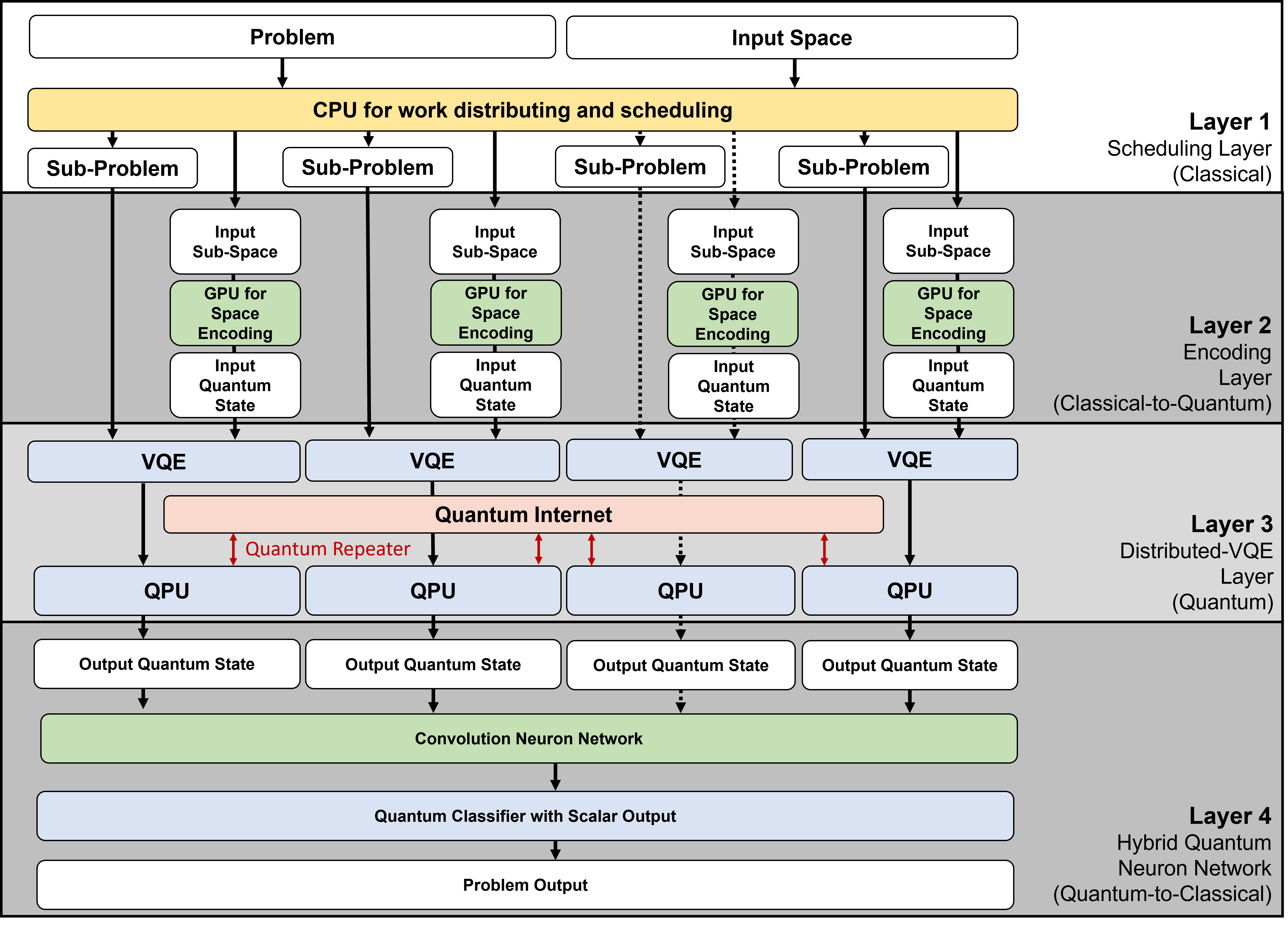}
    \caption{The schematic represents a distributed quantum computing architecture, illustrating the hybrid QCQ framework for classifying phase transitions. This architecture combines CPU-based scheduling with quantum processing across multiple layers and employs GPUs for state encoding. The dashed lines represent omitted \(n\) blocks in parallel due to size constraints of the image}
    \label{fig: Arch}
\end{figure}

To develop a conceptual architecture for distributed quantum computing with a QCQ workflow, we expand upon the concept based on the work of Saurabh et al\cite{saurabh2023conceptual}. This approach enables a deeper integration of quantum computing processes within distributed computing frameworks.
 Based on Fig. \ref{fig: Arch} delineates an integrated architecture designed to leverage distributed quantum resources for advanced computational tasks. The schematic begins with a classical scheduling layer, where a CPU disaggregates a complex problem into sub-problems. These are then converted into quantum states through GPUs in an encoding layer, facilitating the classical-to-quantum transition. The middle of the system can lie in the distributed-VQE layer\cite{parekh2021quantum, khait2023variational}, where QPUs (linked by quantum internet and augmented by quantum repeaters) execute quantum algorithms. Finally, a hybrid quantum-classical neural network analyzes the quantum states, yielding a scalar output via a quantum classifier. This streamlined, multi-layered framework illustrates a scalable approach for executing computationally intensive tasks across distributed quantum systems, aiming to unlock new potentials in near-term quantum information processing.

\section{Result}

\begin{figure}
    \begin{minipage}{0.5\textwidth} 
        \centering 
        \includegraphics[width=\linewidth]{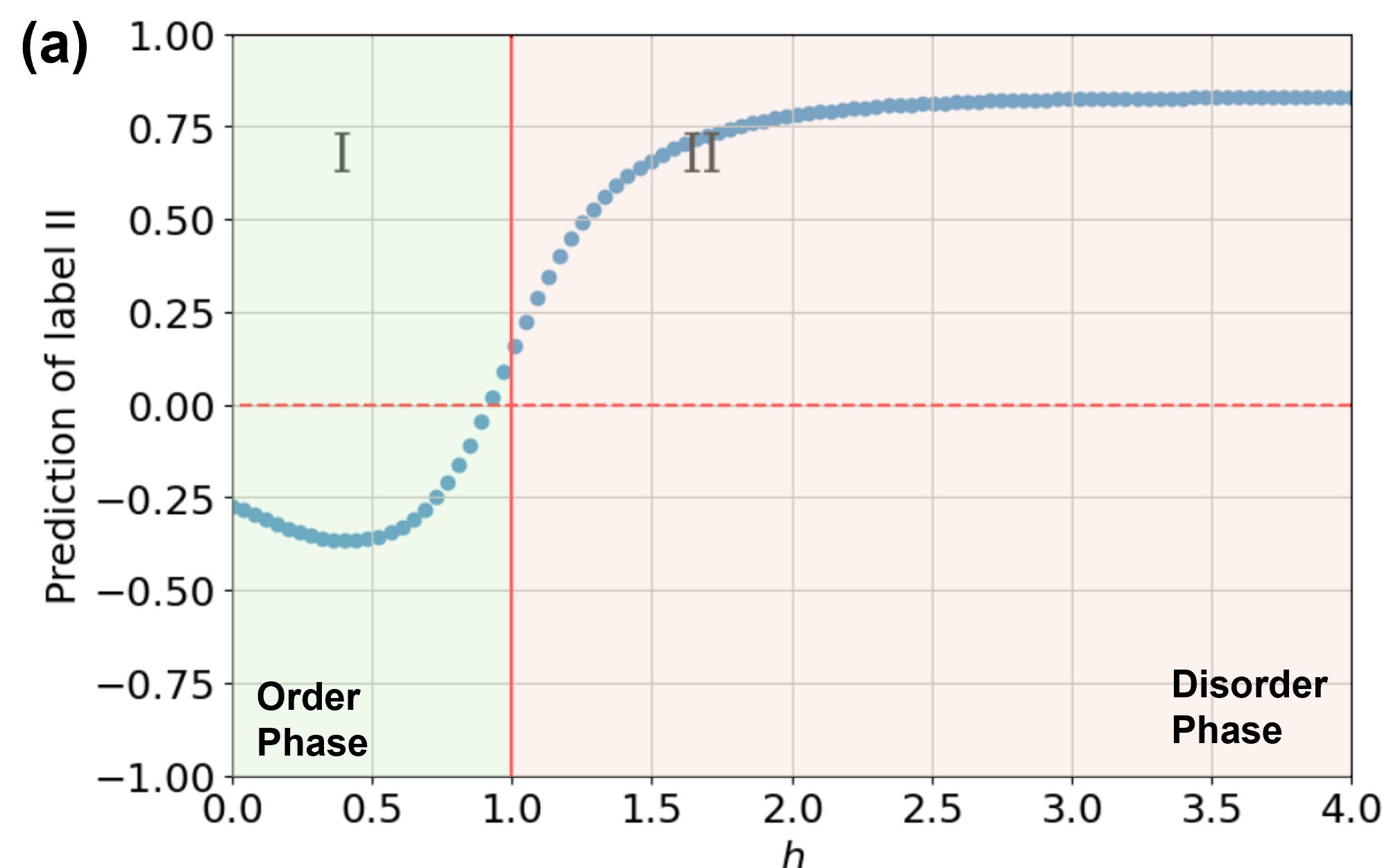}  
    \end{minipage} 
    \begin{minipage}{0.5\textwidth} 
        \centering
        \includegraphics[width=\linewidth]{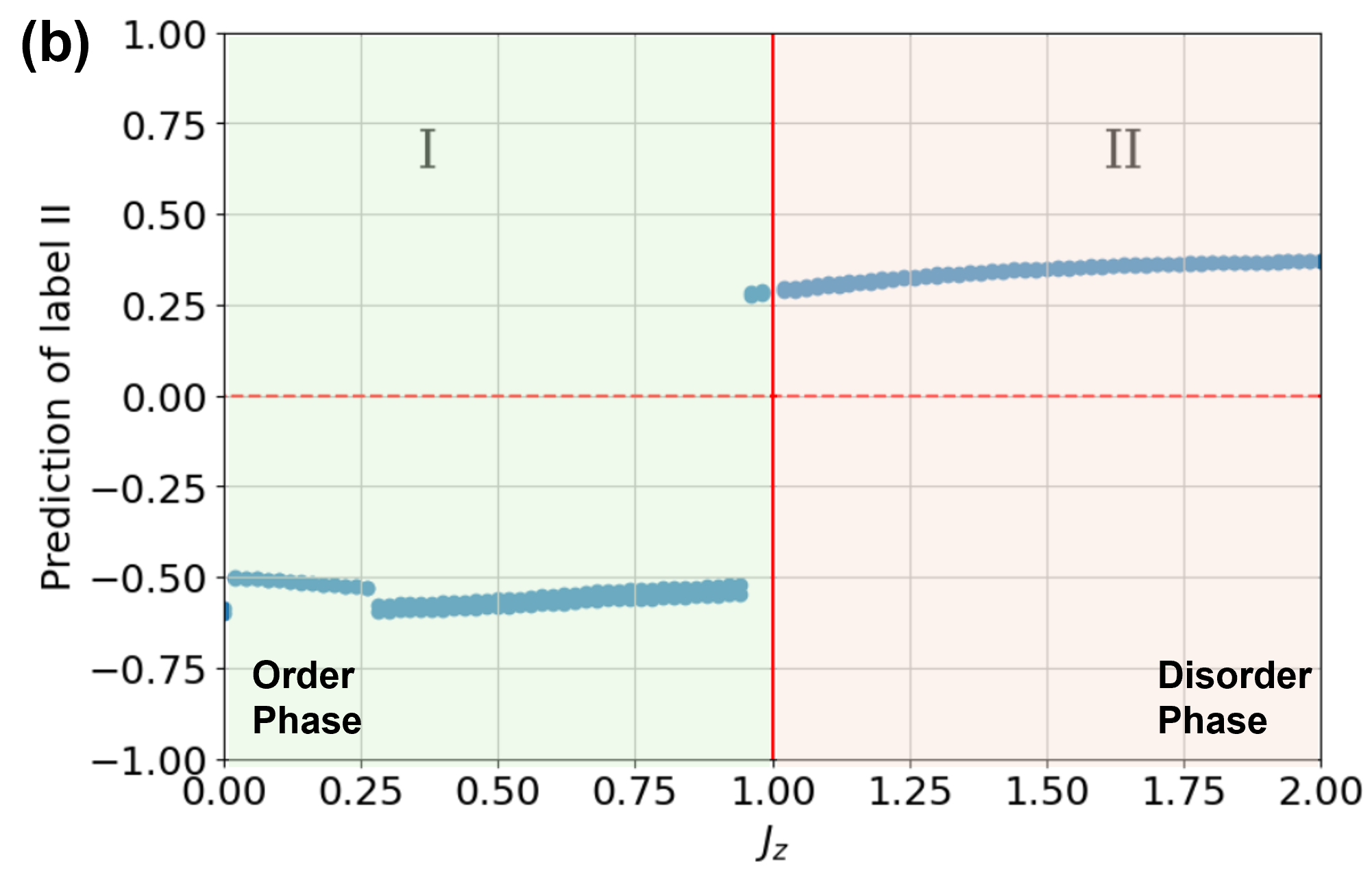} 
    \end{minipage} 
    \caption{(a)Predicted phases as a function of h for the TFIM model. (b) The predicted probability of phases as a function of Jz for the XXZ model. Positive prediction of label II represents phase II, which is above the dashed lines. The theoretically phase II (disorder phase) is the areas on the right-hand side of the red lines (shown in red color).}
    \label{fig: Ising and XXZ phases}
\end{figure}

In Fig. \ref{fig: Ising and XXZ phases}, we have successfully implemented a quantum machine learning classifier for phase transition using the PennyLane package, enhanced by the computational acceleration of the cuQuantum SDK. In our classification task, we observed an increase in accuracy near the phase transition point when utilizing checkerboard states with increasing depth in our VQE algorithm. Notably, even lower-depth ansatz states such as rank-one, tree tensor network, and single-layer checkerboard states delivered comparable results due to the relatively simple nature of the Hamiltonian involved.

\begin{figure}[htpb]
    \centering
    \includegraphics[width=\linewidth]{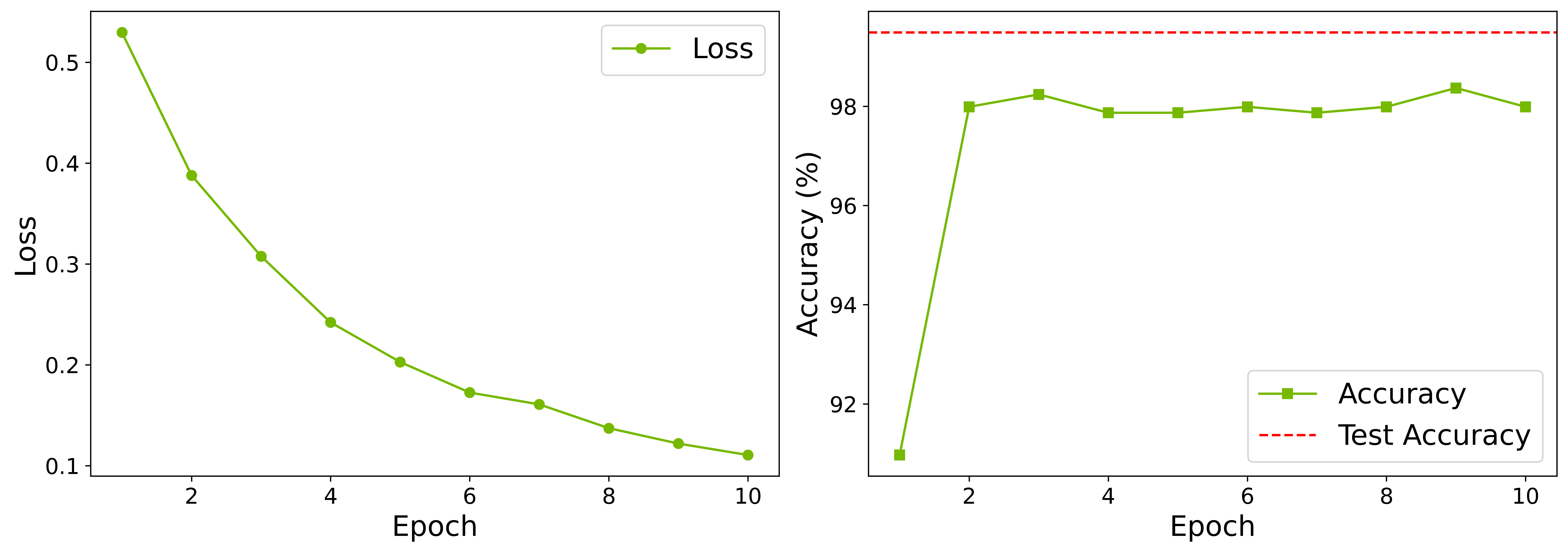}
    \caption{The evolution of loss (left) and accuracy (right) in the QCNN model. As the number of epochs increases, the model's loss decreases exponentially. The model's accuracy improves significantly in the initial epochs and then stabilizes at around 98\%. And test accuracy reaches 99.5\%.}
    \label{fig: QCNN results}
\end{figure}

We prepared a dataset of 100 data points employing a VQE with a four-layered checkerboard ansatz state. After shuffling, this dataset was divided into an 80\% training set and a 20\% test set. Impressively, the classifier achieved a 98.0\% prediction accuracy rate. For the antiferromagnetic XXZ spin chain model, we generated 4000 data points to train the classifier, achieving a 94.6\% accuracy rate on the test data after enhancing the classifier circuit with two additional layers.

The QCNN yields results that are both robust and satisfactory (Fig. \ref{fig: QCNN results}). In the initial phases of the training epochs, the accuracy of the model escalates significantly, demonstrating a rapid learning curve which subsequently reaches a plateau at approximately 98.0\%. Further evaluation on a test set reveals that the model attains a remarkable accuracy of 99.5\%.

The dataset employed for the QCNN model was generated utilizing the VQE on XXZ model. The dataset comprised a total of 1000 data points. In adherence to standard machine learning practices, the dataset was partitioned in an 80-20 split, with 80\% utilized for the training phase and the remaining 20\% reserved for testing.

Our approach demonstrates the potential of quantum-enhanced machine learning to classify phases of matter with high accuracy, outperforming classical Monte-Carlo sampling-based methods. This technique is versatile and can be applied to any model expressible as a spin model, with the capability to be extended to multi-class classification problems. Our results underscore the advantages of integrating quantum simulations with advanced machine learning techniques, paving the way for new insights into the study of quantum phase transitions.

\subsection{GPU Acceleration with cuQuantum for VQE and QCNN}

In this research work, our main task is to develop and optimize a Hybrid QML model with a distributed-quantum-computing framework for classifying phase transitions. In our demonstration, we use the PennyLane package, with a focus on computational efficiency through GPU acceleration with cuQuantum SDK. The inherent challenge in QML is the intensive computational requirement, often necessitating QPUs with high-fidelity qubits. To address this, we employ classical hardware for initial data processing and feature extraction—a step integral to our QML model’s training process. Our strategy utilizes the cuQuantum SDK, which offers a GPU-optimized collection of low-level primitives supporting Pennylane lightning backend. This approach accelerates quantum machine learning model execution on NVIDIA GPUs, substantially shortens training durations, and reduces computational costs. By running our applications on Denvr Dataworks’s CUDA-compatible platform, we benefit from the enhanced processing capabilities of NVIDIA GPUs, which are designed for parallel computing and can therefore handle large datasets and complex operations with greater speed than conventional CPUs. The expected outcome of our work is a threefold acceleration in the training of our quantum machine learning models compared to CPU-only execution. This improvement is attributable to the GPUs’ parallel processing power, which allows for quicker data processing and computation. The integration of cuQuantum with CUDA not only augments computational speed but also offers cost-effective solutions by eliminating the need for expensive quantum hardware and extensive infrastructure. Our objective is to make QML models more scalable and accessible, leveraging these technologies to advance research in quantum phase transitions and establish new performance benchmarks in the field of quantum computing.

\begin{figure}[htpb]
    \centering
    \includegraphics[width=\linewidth]{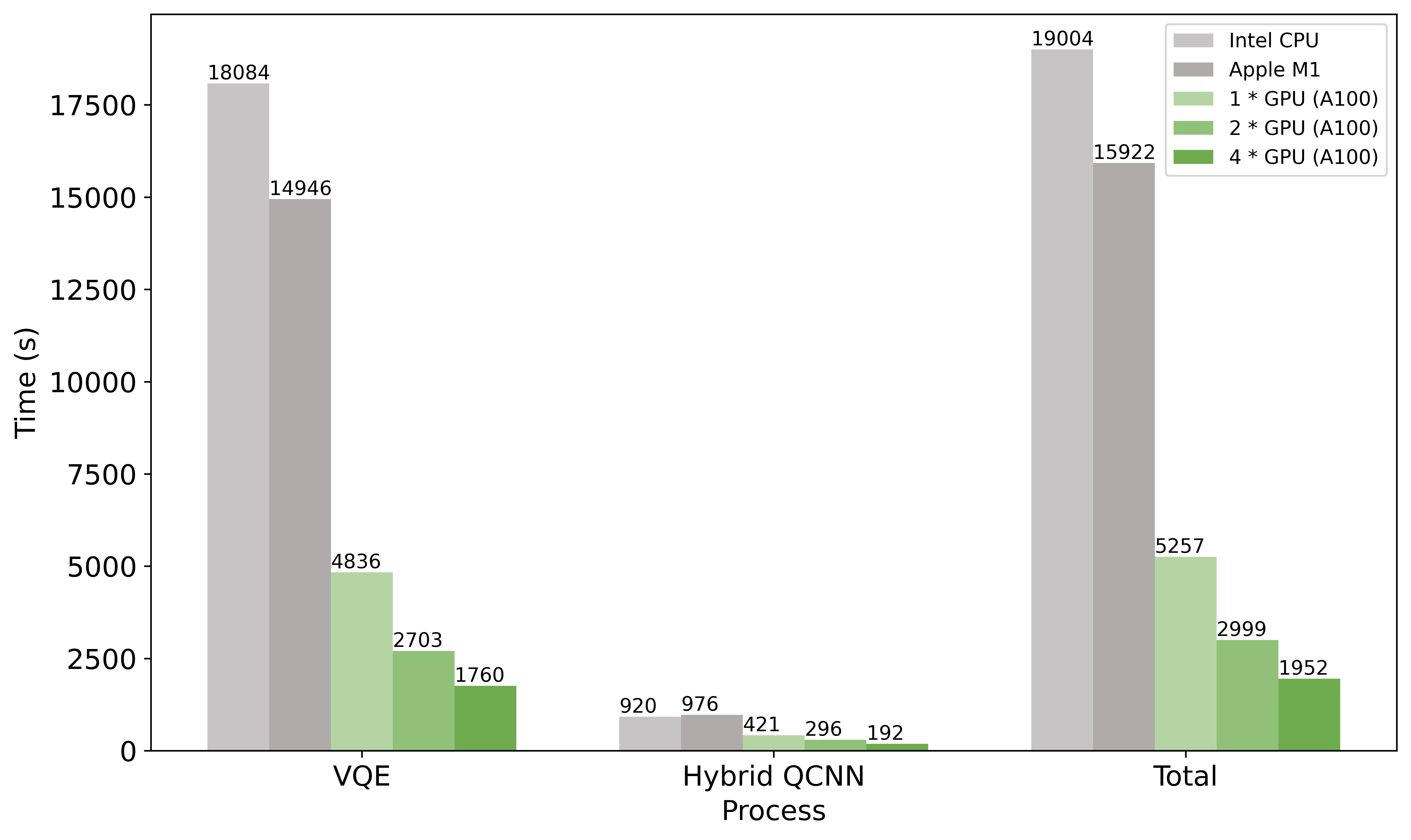}
    \caption{Benchmarking of computation times for VQE, Hybrid QCNN, and Total processes, comparing performance across Intel CPU (i7-13700KF), Apple M1 pro (10-core CPU), and 1, 2, and 4 NVIDIA A100 GPUs as quantum simulators.}
    \label{fig: bemchmarking}
\end{figure}

In line with our objectives, the benchmarking results showcase a comparative analysis of computational performance across different hardware configurations for these quantum simulations. The bar plot distinctly illustrates the time taken to execute specific tasks like the VQE (in this case with 16 qubits), Hybrid QCNN (in this case with 3 qubits), and the aggregate computational process across Intel CPUs, Apple M1 chips, and NVIDIA A100 GPUs. The acceleration impact is significant when utilizing the A100 GPUs, with the most notable performance improvement observed with a configuration of four GPUs, suggesting a near-linear scaling in this particular benchmarking scenario. This benchmarking also agrees with the result shown in Vallero et. al's work\cite{vallero2024state}.

\begin{figure}[htpb]
    \centering
    \includegraphics[width=\linewidth]{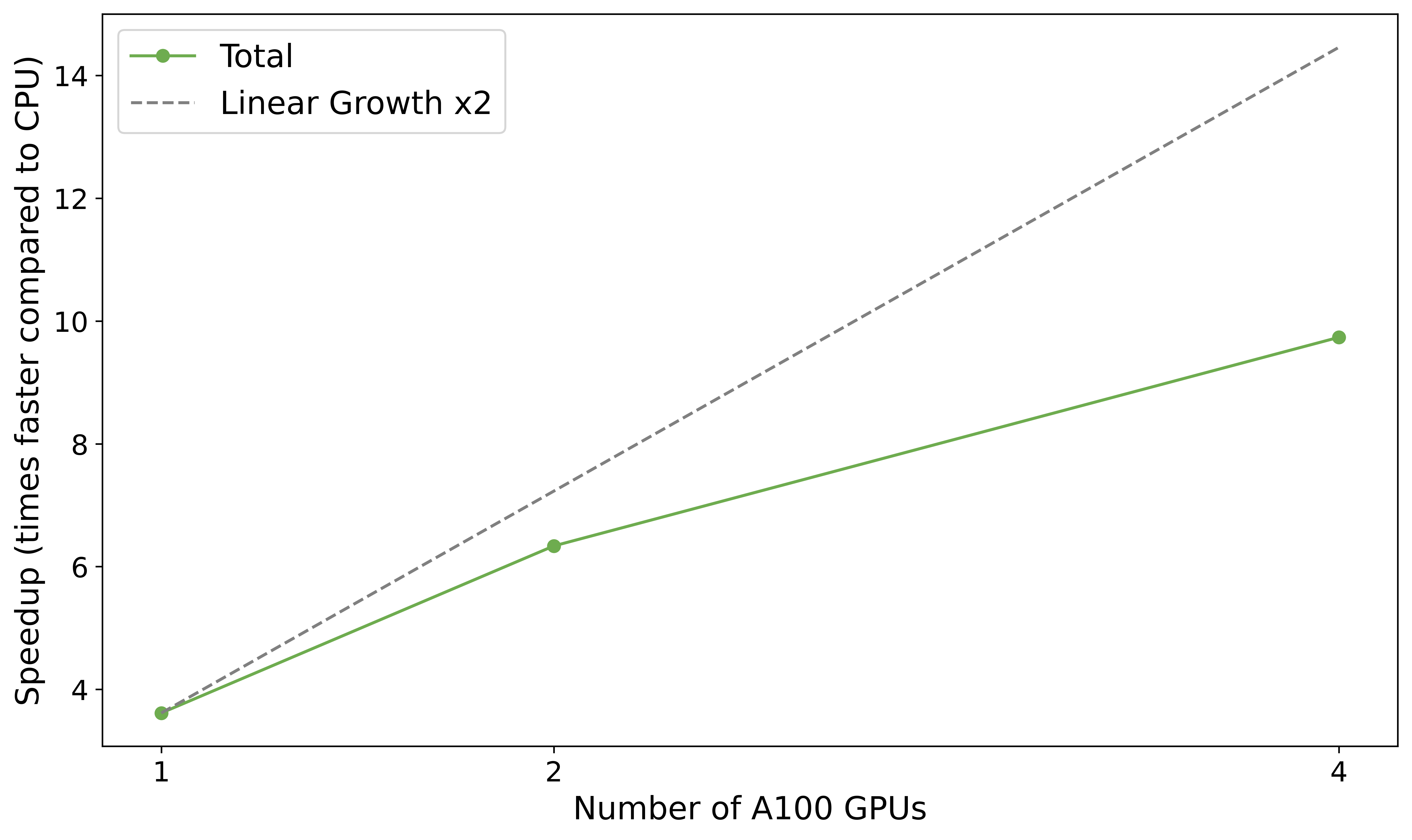}
    \caption{The line graph conducts a linearity performance check of multi-GPU settings for quantum machine learning training.}
    \label{fig: linerity}
\end{figure}

The accompanying line plot further elucidates the speedup achieved with the use of 1, 2, and 4 A100 GPUs compared to CPU benchmarks. It becomes evident that the total speedup gained from GPU acceleration does not fully align with an ideal linear progression, which is visually represented by a grey dashed line. This deviation implies that while GPU acceleration substantially benefits the computational speed, the returns diminish with the addition of more GPUs, possibly due to overhead associated with parallelization and inter-GPU communication. The less efficient may be because in our case the qubit number is not very large. In Vallero et. al's benchmarking\cite{vallero2024state}, with more qubits we run, the more acceleration we will observe. These findings underscore the A100 GPUs' profound capabilities in enhancing performance for quantum simulation tasks, thereby underscoring their potential to facilitate and expedite complex quantum computations in research settings.

\section{Conclusion}

In this study, we propose an innovative QCQ architecture aimed at enhancing applications in distributed quantum computing and Quantum-HPC systems. This architecture integrates VQE algorithms and QCNNs, enabling rapid and accurate quantum simulations, exemplified by phase transition classification within this research. The hybrid QCQ algorithm demonstrated an exceptional 99.5\% test accuracy in predicting phase transitions for models such as the transverse field Ising and XXZ. The architecture efficiently utilizes QPUs with a limited number of high-fidelity qubits, such as superconducting circuits\cite{clarke2008superconducting}, ion traps\cite{kielpinski2002architecture}, neutral atoms\cite{weiss2017quantum}, and NV center diamonds\cite{childress2013diamond}, to scale up the qubit count and achieve distributed quantum computing.

At the heart of the QCQ framework is the seamless integration of variational quantum eigensolver algorithms, tensor network states, and convolutional neural networks. This synergistic amalgamation allows for the effective preparation of quantum states through quantum simulations, followed by their classification using advanced machine learning techniques. The architecture finds applications in quantum reinforcement learning\cite{chen2020variational} and can optimize the hybrid neural network structure using an \(polylog(M)\) approach\cite{liu2024training} or the fast weights approach\cite{yen2024learning}.

The success of this QCQ framework highlights the immense potential of merging quantum computing with machine learning. By surmounting the limitations inherent in classical algorithms, this method paves the way for deeper insights into the behavior of quantum systems across various sizes and configurations. It finds applications in fields such as materials science, condensed matter physics, and sustainability research.

As research in quantum hardware and algorithms progresses, architectures like QCQ will play a critical role in unlocking the transformative capabilities of quantum-enhanced computing. While challenges in further enhancing accuracy and extending applicability remain, the scalability and efficiency exhibited by this framework open up exciting avenues for future exploration. Integrating the strengths of both quantum and classical computing, the QCQ architecture represents a significant advancement towards fully leveraging the potential of quantum technologies in diverse scientific and technological domains.

\section{Future work}
Our upcoming research aims to enhance the QCQ architecture by harnessing the CUDA Quantum platform\cite{kim2023cuda}, delivering substantial speedup and flexibility for quantum-classical computing workflows through distributed quantum simulation and seamless integration with machine learning frameworks.

\section*{Acknowledgment}
The authors extend their gratitude to Andreas Hehn, Tobi Odufeso, and Tom Lubowe for invaluable assistance and insights, which have been pivotal to the success of this research work. This research was supported by Xanadu, NVIDIA, Amazon, and Denvr Dataworks under QHack 2024. K.C. is grateful for the financial support from both the Turing Scheme for the Imperial Global Fellows Fund and the Imperial QuEST Seed Fund. Special acknowledgement goes to the Quantum Open Source Foundation (QOSF) and QuantumPedia AI for their contributions and their quantum computing mentorship program.

\section*{Code Avalibility}
The source code for generating the dataset and figures presented in this research work is openly accessible at \url{https://github.com/Louisanity/cuPhastLearn}.

\bibliographystyle{siamurl}
\bibliography{mybib}

\begin{thebibliography}{10}

\bibitem{biamonte2017quantum}
{\sc J.~Biamonte, P.~Wittek, N.~Pancotti, P.~Rebentrost, N.~Wiebe and S.~Lloyd}, {\em Quantum machine learning}, Nature {\bfseries 549} (2017), pp.~195--202.

\bibitem{benedetti2019parameterized}
{\sc M.~Benedetti, E.~Lloyd, S.~Sack and M.~Fiorentini}, {\em Parameterized quantum circuits as machine learning models}, Quantum Science and Technology {\bfseries 4} (2019), p.~043001.

\bibitem{ngairangbam2022anomaly}
{\sc V.~S. Ngairangbam, M.~Spannowsky and M.~Takeuchi}, {\em Anomaly detection in high-energy physics using a quantum autoencoder}, Physical Review D {\bfseries 105} (2022), p.~095004.

\bibitem{zoufal2019quantum}
{\sc C.~Zoufal, A.~Lucchi and S.~Woerner}, {\em Quantum generative adversarial networks for learning and loading random distributions}, npj Quantum Information {\bfseries 5} (2019), p.~103.

\bibitem{mari2020transfer}
{\sc A.~Mari, T.~R. Bromley, J.~Izaac, M.~Schuld and N.~Killoran}, {\em Transfer learning in hybrid classical-quantum neural networks}, Quantum {\bfseries 4} (2020), p.~340.

\bibitem{chen2022quantum}
{\sc S.~Y.-C. Chen, T.-C. Wei, C.~Zhang, H.~Yu and S.~Yoo}, {\em Quantum convolutional neural networks for high energy physics data analysis}, Physical Review Research {\bfseries 4} (2022), p.~013231.

\bibitem{abbas2021power}
{\sc A.~Abbas, D.~Sutter, C.~Zoufal, A.~Lucchi, A.~Figalli and S.~Woerner}, {\em The power of quantum neural networks}, Nature Computational Science {\bfseries 1} (2021), pp.~403--409.

\bibitem{chen2023quantum}
{\sc K.-C. Chen, X.~Xu, H.~Makhanov, H.-H. Chung and C.-Y. Liu}, {\em Quantum-enhanced support vector machine for large-scale stellar classification with gpu acceleration}, arXiv preprint arXiv:2311.12328  (2023).

\bibitem{marshall2023high}
{\sc S.~C. Marshall, C.~Gyurik and V.~Dunjko}, {\em High dimensional quantum machine learning with small quantum computers}, Quantum {\bfseries 7} (2023), p.~1078.

\bibitem{mundada2023experimental}
{\sc P.~S. Mundada et~al.}, {\em Experimental benchmarking of an automated deterministic error-suppression workflow for quantum algorithms}, Physical Review Applied {\bfseries 20} (2023), p.~024034.

\bibitem{chen2023short}
{\sc K.-C. Chen}, {\em Short-depth circuits and error mitigation for large-scale ghz-state preparation, and benchmarking on ibm's 127-qubit system}, in 2023 IEEE International Conference on Quantum Computing and Engineering (QCE), vol.~2, IEEE, 2023, pp.~207--210.

\bibitem{huang2022quantum}
{\sc H.-Y. Huang et~al.}, {\em Quantum advantage in learning from experiments}, Science {\bfseries 376} (2022), pp.~1182--1186.

\bibitem{monaco2023quantum}
{\sc S.~Monaco, O.~Kiss, A.~Mandarino, S.~Vallecorsa and M.~Grossi}, {\em Quantum phase detection generalization from marginal quantum neural network models}, Physical Review B {\bfseries 107} (2023), p.~L081105.

\bibitem{grzesiak2020efficient}
{\sc N.~Grzesiak et~al.}, {\em Efficient arbitrary simultaneously entangling gates on a trapped-ion quantum computer}, Nature communications {\bfseries 11} (2020), p.~2963.

\bibitem{cuomo2020towards}
{\sc D.~Cuomo, M.~Caleffi and A.~S. Cacciapuoti}, {\em Towards a distributed quantum computing ecosystem}, IET Quantum Communication {\bfseries 1} (2020), pp.~3--8.

\bibitem{alexeev2023quantum}
{\sc Y.~Alexeev et~al.}, {\em Quantum-centric supercomputing for materials science: A perspective on challenges and future directions}, arXiv preprint arXiv:2312.09733  (2023).

\bibitem{awschalom2021development}
{\sc D.~Awschalom et~al.}, {\em Development of quantum interconnects (quics) for next-generation information technologies}, PRX Quantum {\bfseries 2} (2021), p.~017002.

\bibitem{prest2023quantum}
{\sc M.~Prest and K.-C. Chen}, {\em Quantum-error-mitigated detectable byzantine agreement with dynamical decoupling for distributed quantum computing}, arXiv preprint arXiv:2311.03097  (2023).

\bibitem{saurabh2023conceptual}
{\sc N.~Saurabh, S.~Jha and A.~Luckow}, {\em A conceptual architecture for a quantum-hpc middleware}, in 2023 IEEE International Conference on Quantum Software (QSW), IEEE, 2023, pp.~116--127.

\bibitem{cuQuantum}
{\sc NVIDIA}, {\em {cuQuantum SDK}: Simulating quantum circuits on {GPU}s}.
\newblock \url{https://developer.nvidia.com/cuquantum-sdk}, 2021.
\newblock Accessed: February 23, 2023.

\bibitem{bergholm2018pennylane}
{\sc V.~Bergholm et~al.}, {\em Pennylane: Automatic differentiation of hybrid quantum-classical computations}, arXiv preprint arXiv:1811.04968  (2018).

\bibitem{gao2021scaling}
{\sc H.~Gao and N.~Sakharnykh}, {\em Scaling joins to a thousand gpus.}, in ADMS@ VLDB, 2021, pp.~55--64.

\bibitem{wintersperger2022qpu}
{\sc K.~Wintersperger, H.~Safi and W.~Mauerer}, {\em Qpu-system co-design for quantum hpc accelerators}, in International Conference on Architecture of Computing Systems, Springer, 2022, pp.~100--114.

\bibitem{uvarov2020machine}
{\sc A.~Uvarov, A.~Kardashin and J.~D. Biamonte}, {\em Machine learning phase transitions with a quantum processor}, Physical Review A {\bfseries 102} (2020), p.~012415.

\bibitem{khait2023variational}
{\sc I.~Khait, E.~Tham, D.~Segal and A.~Brodutch}, {\em Variational quantum eigensolvers in the era of distributed quantum computers}, arXiv preprint arXiv:2302.14067  (2023).

\bibitem{liu2023learning}
{\sc C.-Y. Liu, C.-H.~A. Lin and K.-C. Chen}, {\em Learning quantum phase estimation by variational quantum circuits}, arXiv preprint arXiv:2311.04690  (2023).

\bibitem{diadamo2021distributed}
{\sc S.~DiAdamo, M.~Ghibaudi and J.~Cruise}, {\em Distributed quantum computing and network control for accelerated vqe}, IEEE Transactions on Quantum Engineering {\bfseries 2} (2021), pp.~1--21.

\bibitem{du2022distributed}
{\sc Y.~Du, Y.~Qian, X.~Wu and D.~Tao}, {\em A distributed learning scheme for variational quantum algorithms}, IEEE Transactions on Quantum Engineering {\bfseries 3} (2022), pp.~1--16.

\bibitem{modani2023multi}
{\sc M.~Modani, A.~Banerjee and A.~Das}, {\em Multi-gpu-enabled quantum circuit simulations on hpc-ai system}, in International Conference on Data Management, Analytics \& Innovation, Springer, 2023, pp.~845--854.

\bibitem{liu2019variational}
{\sc J.-G. Liu, Y.-H. Zhang, Y.~Wan and L.~Wang}, {\em Variational quantum eigensolver with fewer qubits}, Physical Review Research {\bfseries 1} (2019), p.~023025.

\bibitem{Utkarsh2023Spin}
{\sc D.~G. Utkarsh~Azad}, {\em Pennylane spin datasets}.
\newblock \url{https://pennylane.ai/datasets/qspin/ising-model}, 2023.

\bibitem{huang2020predicting}
{\sc H.-Y. Huang, R.~Kueng and J.~Preskill}, {\em Predicting many properties of a quantum system from very few measurements}, Nature Physics {\bfseries 16} (2020), pp.~1050--1057.

\bibitem{cong2019quantum}
{\sc I.~Cong, S.~Choi and M.~D. Lukin}, {\em Quantum convolutional neural networks}, Nature Physics {\bfseries 15} (2019), pp.~1273--1278.

\bibitem{xxzphase}
{\sc F.~Franchini}, {\em An Introduction to Integrable Techniques for One-Dimensional Quantum Systems}, Springer International Publishing, 2017.
\newblock {\sc url: }\url{http://dx.doi.org/10.1007/978-3-319-48487-7}, \href {http://dx.doi.org/10.1007/978-3-319-48487-7} {\path{doi:10.1007/978-3-319-48487-7}}.

\bibitem{parekh2021quantum}
{\sc R.~Parekh, A.~Ricciardi, A.~Darwish and S.~DiAdamo}, {\em Quantum algorithms and simulation for parallel and distributed quantum computing}, in 2021 IEEE/ACM Second International Workshop on Quantum Computing Software (QCS), IEEE, 2021, pp.~9--19.

\bibitem{vallero2024state}
{\sc M.~Vallero, F.~Vella and P.~Rech}, {\em State of practice: evaluating gpu performance of state vector and tensor network methods}, arXiv preprint arXiv:2401.06188  (2024).

\bibitem{clarke2008superconducting}
{\sc J.~Clarke and F.~K. Wilhelm}, {\em Superconducting quantum bits}, Nature {\bfseries 453} (2008), pp.~1031--1042.

\bibitem{kielpinski2002architecture}
{\sc D.~Kielpinski, C.~Monroe and D.~J. Wineland}, {\em Architecture for a large-scale ion-trap quantum computer}, Nature {\bfseries 417} (2002), pp.~709--711.

\bibitem{weiss2017quantum}
{\sc D.~S. Weiss and M.~Saffman}, {\em Quantum computing with neutral atoms}, Physics Today {\bfseries 70} (2017), pp.~44--50.

\bibitem{childress2013diamond}
{\sc L.~Childress and R.~Hanson}, {\em Diamond nv centers for quantum computing and quantum networks}, MRS bulletin {\bfseries 38} (2013), pp.~134--138.

\bibitem{chen2020variational}
{\sc S.~Y.-C. Chen, C.-H.~H. Yang, J.~Qi, P.-Y. Chen, X.~Ma and H.-S. Goan}, {\em Variational quantum circuits for deep reinforcement learning}, IEEE Access {\bfseries 8} (2020), pp.~141007--141024.

\bibitem{liu2024training}
{\sc C.-Y. Liu et~al.}, {\em Training classical neural networks by quantum machine learning}, arXiv preprint arXiv:2402.16465  (2024).

\bibitem{yen2024learning}
{\sc S.~Yen-Chi~Chen}, {\em Learning to program variational quantum circuits with fast weights}, arXiv e-prints  (2024), pp.~arXiv--2402.

\bibitem{kim2023cuda}
{\sc J.-S. Kim, A.~McCaskey, B.~Heim, M.~Modani, S.~Stanwyck and T.~Costa}, {\em Cuda quantum: The platform for integrated quantum-classical computing}, in 2023 60th ACM/IEEE Design Automation Conference (DAC), IEEE, 2023, pp.~1--4.

\end{thebibliography}

\end{document}